# Chiral Assemblies of Pinwheel Superlattices on Substrates

Shan Zhou,[1,2,13] Jiahui Li,[1,13] Jun Lu,[3,4,13] Haihua Liu,[5] Ji-Young Kim,[3,4] Ahyoung Kim,[1] Lehan Yao,[1] Chang Liu,[1] Chang Qian,[1] Zachary D. Hood,[6] Wenxiang Chen,[1,2] Thomas E. Gage,[5] Ilke Arslan,[5] Alex Travesset,[7,8] Kai Sun,[9] Nicholas A. Kotov,[3,4,10*] and Qian Chen[1,2,11,12*]

[1]Department of Materials Science and Engineering, [2]Materials Research Laboratory, [11]Beckman Institute for Advanced Science and Technology, [12]Department of Chemistry, University of Illinois at Urbana-Champaign, Urbana, Illinois 61801, United States

[3]Department of Chemical Engineering, [4]Biointerfaces Institute, [9]Department of Physics, [10]Department of Materials Science and Engineering, University of Michigan, Ann Arbor, Michigan 48109, United States

[5]Center for Nanoscale Materials, [6]Applied Materials Division, Argonne National Laboratory, Lemont, Illinois 60439, United States

[7]Department of Physics and Astronomy, [8]Department of Materials Science and Engineering, Iowa State University and Ames Lab, Ames, Iowa 50011, United States

*To whom correspondence should be addressed: kotov@umich.edu; qchen20@illinois.edu

[13]These authors contributed equally to this work.

**The unique topology and physics of chiral superlattices make their self-assembly from nanoparticles a holy grail for (meta)materials[1-3]. Here we show that tetrahedral gold nanoparticles can spontaneously transform from a perovskite-like low-density phase with corner-to-corner connections into pinwheel assemblies with corner-to-edge connections and denser packing. While the corner-sharing assemblies are achiral, pinwheel superlattices become strongly mirror-asymmetric on solid substrates as demonstrated by chirality measures. Liquid-phase transmission electron microscopy and computational models show that van der Waals and electrostatic interactions between nanoparticles control thermodynamic equilibrium. Variable corner-to-edge connections among tetrahedra enable fine-tuning of chirality. The domains of the bilayer superlattices display strong chiroptical activity identified by photon-induced near-field electron microscopy and finite-difference time-domain simulations. The simplicity and versatility of the substrate-supported chiral superlattices facilitate manufacturing of metastructured coatings with unusual optical, mechanical and electronic characteristics.**

Superlattices self-assembled from nanoparticles (NPs) have attracted extensive research attention due to their versatile composition and packing-dependent optical, magnetic, electronic, catalytic and mechanical properties[1-6]. They have been made from a variety of semiconductor, metal and ceramic materials displaying a wide range of packing motifs, yet, the experimental assembly of chiral NP superlattices remains fundamentally difficult[7,8]. Imparting chirality to NP superlattices and their thin films is needed for technological advances of (meta)surfaces because of the unique mechanical deformations and optical effects possible in chiral nanoassemblies[9,10]. Spontaneous thermodynamics-driven self-assembly of NPs is fundamentally challenging because constituent particles simultaneously require exceptional uniformity and high asymmetry, which requires new methods of enantioselective synthesis vastly different from previous chemical methodologies[11,12]. The strong face-to-face attraction between NPs also favours closely packed achiral superlattices[7,8,13]. Templated assemblies are capable of circumventing this challenge by assembling achiral NPs with biomolecular ligands on helical scaffolds that have proven to be effective for translating chirality into clusters of a few particles[14,15] but not into large arrays; NPs connected by biological ligands produced colloidal crystals with achiral packings[16,17].

Learning from condensed matter physics, it should be possible to self-assemble open chiral superlattices from NPs by transforming non-closely packed achiral corner-sharing structures into new states with higher packing fraction by reconfiguring building blocks with tetrahedral and other polyhedral geometries (Fig. 1). No chirality of the constitutive particles is needed. In fact, numerous achiral ones can be utilized. Some prominent examples of corner-sharing materials (Fig. 1a) include perovskite, pyrochlore, diamond and zeolite, which are widely explored to impart ferroelectricity, piezoelectricity, superconductivity, spin filtering, optical activity and other properties to different materials[18,19]. In these materials, the symmetry-breaking phase transitions can be driven by changes in pressure, stoichiometry, temperature, or the surface chemistry of the building blocks[20-24].

The same geometrical advantages that make the assembly of open chiral superlattices from tetrahedral building blocks attractive, however, also hinder their experimentally realization. It is well known that tetrahedra can produce singular helices or quasi-crystals but self-assembly into regular periodic superlattices encounters a large number of kinetically arrested states[7,8,25,26]. In this paper, we show how the set of multifaceted difficulties mentioned above, is overcome starting from a robust method to synthesize and purify gold tetrahedral NPs with a tunable edge length $L$ with low dispersity (Fig. 2a, Methods, Supplementary Figs. 1,2, Supplementary Table 1). Compared with previous studies on tetrahedral semiconductor NPs that assemble face-to-face[7,8,25], gold tetrahedra are mostly coated by charged ligands (Supplementary Fig. 3, Supplementary Table 2), which increases the face-to-face repulsion.

Controlled evaporation of an aqueous solution of nanoscale gold tetrahedra on Si wafer drives their assembly into nontrivial bilayer lattices composed of two interlacing sets of corner-sharing tetrahedra (Fig. 1b, Fig. 2a–d, Methods). It has a low packing fraction of 1/3 (Fig. 1c), lower than those of all the predicted assembly structures of tetrahedra[27] (Supplementary Fig. 4, Supplementary Note 1). As shown in Fig. 2d, the bottom-layer tetrahedra sit on the Si substrate with corners pointing up (orange triangles), and top-layer tetrahedra fill the triangular voids with corners pointing down (blue triangles). Fast Fourier transform (FFT) of the scanning electron microscopy (SEM) image shows a pattern resembling a honeycomb structure (Fig. 2d). The unit cell of the bilayer lattice shows a pair of neighbouring tetrahedral NPs facing opposite directions along the $z$-axis (Fig. 2b). Particle tracking measures a large face-to-face distance $D$ of 11.2 nm between this NP pair (Fig. 2c, Supplementary Figs. 5–7, Supplementary Notes 1,2); even considering the thickness of adsorbed ligands (3.2 nm on each tetrahedron) [28], this NP pair is not physically in contact at the faces (Fig. 2c, Supplementary Table 3). Our coarse-grained modelling of the total pairwise interparticle interaction shows that the achiral bilayer lattice with nearly perfect corner-to-corner connection is favoured for small-sized gold tetrahedra at a low ionic strength $I$ (Fig. 1d, Supplementary Note 3) by balancing van der Waals attraction ($E_{vdW}$) between tetrahedra and electrostatic repulsion ($E_{el}$) between charged ligands, which can be achieved in the solution during the drying process. The strong face-to-face repulsion ($0.59k_BT$) separates the faces of tetrahedra apart, allowing only corner-to-corner connections, leading to this previously inaccessible low packing fraction lattice.

Pinwheel lattices with corner-to-edge connections and greater packing fractions are achieved when a spontaneous in-plane "compression" is induced by weakening $E_{el}$ and/or increasing $E_{vdW}$. For example, large tetrahedra increase $E_{vdW}$, and hence, form extended domains (Supplementary Fig. 8) of pinwheel packing (Fig. 2e) with corner-to-edge connections, where decreased $D$ and increased packing fraction are observed (Fig. 1c). This twisted pinwheel structure is not predicted by existing theory or simulations[26,27,29-31] and has not been realized experimentally for any nanoscale particles[7,8,32]. Note that only in the bilayer superlattice all tetrahedra slide into the same direction and to the same degree due to the equilateral constraints imposed by the top layer of tetrahedra, which would not be the case for a monolayer of tetrahedral particles (Fig. 1b, Supplementary Video 1). The corresponding FFT pattern exhibits a consistent twist over a hexagonal pattern characteristic of the original achiral lattice (Fig. 2e). The effective "compression" increases the packing fraction to 2/3 when all tetrahedra rotate in-plane by 30° (Fig. 1b,c, Supplementary Note 1) and changes the point group from $D_{3d}$ to $S_6$ (Fig. 2c, Supplementary Video 1).

We note that free-standing pinwheel superlattices of the corner-to-edge assemblies of tetrahedra are inversion symmetric, but they become chiral when deposited on a flat substrate. The pinwheels can be superimposed onto their mirror image if flipping in $z$ direction is allowed, but when the substrate is present, the up and down sides are no longer the same so flipping results into an inequivalent mirror image. This effect can be demonstrated by calculating the chirality measure with the Osipov-Pickup-Dunmur parameter[33,34] (OPD) of a freestanding pinwheel lattice in the absence of a substrate (OPD = 0) (Supplementary Fig. 9, Supplementary Note 4, Supplementary Table 5), which becomes either positive or negative indicating the assembly of left and right chirality with inclusion of a substrate (Fig. 3a, Supplementary Fig. 10, Supplementary Table 6). This substrate effect is captured in the OPD calculation by considering not only the centre of mass position of each tetrahedron, but also those of corners and substrate (Supplementary Note 4). |OPD| of the total structure changes from 0 to higher value when the corner offset ($\Delta$, see definition in Fig. 1c and Supplementary Note 2), a physical measurement of symmetry breaking, increases (Fig. 3a, Supplementary Table 6). Experimentally the pinwheel superlattices must always sit on the substrate, making the domains with opposite handedness truly mirror-asymmetric enantiomers. Incident polarized light thus distinguishes the top and bottom tetrahedra and undergoes polarization rotation, making the pinwheels chiroptically active. This chirality can be enumerated by the corner offset $\Delta$. Variation of NP size, ionic strength and molecular additives results in systematic changes of |$\Delta$| from 0.03 to 0.56 (Supplementary Figs. 5,6, Supplementary Table 4). Correspondingly, |OPD| for the lattices on the substrate changes roughly from 0 to 1.68.

The structure-dependent chiroptical properties of these superlattices are demonstrated by their optical activity measured using photon-induced near-field electron microscopy (PINEM) [35-37] in an ultrafast electron microscope (UEM) (Fig. 3, Supplementary Figs. 11–19). PINEM maps plasmonic fields near the nanostructures by selecting electrons that gain energy due to photon–electron interactions when electrons overlap photons spatiotemporally near the nanostructures (Fig. 3b). This method has been utilized for studies of achiral nanostructures[38], but it can also be utilized for chiral nanostructures, when the traditional macroscale circular dichroism (CD) spectroscopy is inapplicable. We use left-handed (LCP) and right-handed circularly polarized (RCP) light to excite the plasmonic fields and monitor chiral response. Compared with state-of-the-art methods of single particle CD or polarized light optical microscopy[39], PINEM measures directly the electric field (*E*-field) distributions of nanostructures with nanometre spatial resolution (i.e., two orders of magnitude higher).

Pump laser pulses with LCP or RCP light polarizations illuminate the pinwheel assemblies with different $\Delta$, handedness, and domain sizes (Fig. 3b). For the chiral pinwheel lattices, *E*-field distributions have apparent differences under the irradiation of LCP and RCP (Fig. 3b–f, Supplementary Figs. 11,12). With the four-particle domain of $\Delta = -0.2$ as an example, strong plasmonic resonance in the *E*-field is located at the right corner under LCP, in comparison to that located at the top corner under RCP (Fig. 3b), consistent with the *E*-field maps obtained from finite-difference time-domain (FDTD) simulation (Supplementary Figs. 13,14). The optical asymmetry can be further highlighted *via* the subtraction of *E*-field maps under LCP and RCP (Fig. 3b, Supplementary Fig. 12, Supplementary Note 5) with outcomes matching nearly perfectly the FDTD simulations (Supplementary Figs. 13–15). The calculated CD spectra exhibit redshifts of peaks and increase in intensity with the increase of domain sizes (particle number from 4 to 73) (Fig. 3d,f), indicating strong optical asymmetry for these metamaterials.

At similar domain sizes, the chiroptical activity shows a clear dependency on $\Delta$ (Fig. 3c–e). For the domain with $\Delta = 0$, the subtracted *E*-field has a weak and alternative localized pattern, which corresponds to nearly zero signal in the CD spectra (Supplementary Fig. 16). For the domain with a limited $\Delta = 0.1$, the subtracted *E*-field distribution becomes diffusive around the domain in experiments, and shows a similar weak chiral localized feature and low CD peak in the simulation (Fig. 3e). For domains with a large $|\Delta| = 0.2$–0.4, the subtracted *E*-field distribution adopts a pattern with clearly asymmetrical localized and alternative signals with opposite signs outside the domain (Fig. 3c,d). The normalized circular profile for the domain with higher $|\Delta|$ shows more distinct features (Fig. 3c–e), which also matches with the appearance of greater amplitude in the calculated CD spectra (Supplementary Figs. 17,18). The domains with opposite handedness exhibit reversed peaks in CD responses (Fig. 3c,d), indicating that the pinwheel lattice represents the true mirror asymmetric enantiomers.

The structural dependence of the chiroptical activity enables tunable symmetry breaking. Our control evaporation experiments at low particle concentrations (Fig. 4a,b, Supplementary Fig. 20a) point to the mechanism that individual tetrahedral NPs first assemble into $N = 4$ pinwheel (the smallest to adopt chirality), which grows into extended domains with identical handedness. This process is corroborated by liquid-phase transmission electron microscopy (TEM) experiments[40-43] (Fig. 4c, Methods, Supplementary Note 6), where individual tetrahedral NPs are observed to assemble and extend into a bilayer lattice within seconds (Fig. 4d, Supplementary Fig. 20b) upon an increase in ionic strength *I*. Thus, we find that the sum of one interlayer interaction and two intralayer interactions is able to predict the stable pinwheel structure matching quantitatively the experiments (Supplementary Note 3, Supplementary Figs. 21–23). The predicted energy minimum locates at increasing $\Delta$ as the particle $L$ increases (Fig. 4e), consistent with our experimentally observed $\Delta$ increase (Fig. 4g). The predicted total energy minimum also occurs at an increasing when $I$ increases (Fig. 4f, Supplementary Fig. 21), which we confirm in experiment (Fig. 4g). As we further increase $I$ to "compress" the open superlattices more, the symmetry breaking can be extended into three dimensions (Fig. 4h,i). Multilayer structures with corner protrusion in all the top bilayers (Fig. 4h) or tilted assemblies without three-fold symmetry (Fig. 4i) are formed, rendering them truly chiral even in the absence of substrate. These assemblies with variable offsets or tilts around $z$-axis can further decrease $D$ and increase packing fraction. An even higher $I$ leads to a complete collapse into dense face-to-face decahedral packing (Fig. 4j).

**Conclusion**. Previously unknown perovskite-like corner-sharing superlattices with ultralow packing density were self-assembled from nanoscale tetrahedra. The equilibrium configurations of the pinwheel structural motif in these superlattices can be controlled by attractive and repulsive interactions between NPs, leading to superlattices with variable chirality. While the original corner-sharing nanoassemblies are achiral, the compacted superlattices with corner-to-edge connections become chiral when resting on solid substrates. These findings can be extended to a large library of achiral particles, their non-closely packed assemblies, and lattice reconfiguration pathways, which also allow for rich design space of metastructured surfaces with local-specific chiroptical activity. Besides unique polarization rotation, these superlattices are expected to exhibit an emergent conformal symmetry (Supplementary Fig. 24), and zero-energy edge modes, in analogy to the twisted kagome lattice[44,45].

**Acknowledgements**

Experiments were carried out in part in the Materials Research Laboratory (MRL) Central Research Facilities, University of Illinois. We thank Dr. Jessica Spear and Dr. Honghui Zhou (MRL, University of Illinois) for assistance with SEM measurements.

**Author contributions**

S.Z., J.Li, J.Lu and Q.C. designed the experiments. S.Z. and J.Li performed the experiments and data analysis. J.Li performed interaction calculation. J.Lu and N.A.K. performed FDTD simulation and analysed simulation data. H.L., Z.D.H., T.E.G. and I.A. performed PINEM experiments. J-Y.K. performed OPD calculation. S.Z. and A.K. performed tetrahedra purification. L.Y. developed protocols for image analysis and particle tracking. C.L., C.Q. and J.Li performed liquid-phase TEM experiments. W.C. contributed to property discussion. A.T. contributed to geometric calculation and analysis. K.S. contributed to Maxwell lattice construction and discussion. All authors contributed to the writing of the paper. Q.C. and N.A.K. supervised the work.

**Funding**

This research was supported by the Office of Naval Research (MURI N00014-20-1-2479). S.Z. and Q.C. thank the support from Alfred Sloan Foundation for the Sloan fellowship. The authors also acknowledge the Center for Nanoscale Materials, a U.S. Department of Energy Office of Science User Facility, supported by the U.S. Department of Energy, Office of Basic Energy Sciences, under Contract No. DE-AC02-06CH11357. N.A.K., J. Lu, and J.-Y.K. thank Vannevar Bush DoD Fellowship to N.A.K. titled "Engineered Chiral Ceramics" ONR N000141812876. The FDTD simulation and OPD calculation were supported by NSF 1463474 titled "Energy- and Cost-Efficient Manufacturing Employing Nanoparticles". A.T. thanks the support by NSF DMR-CMMT 1606336 on NP packing analysis.

**Competing interests**

The authors declare no competing interests.

**Correspondence and requests for materials** should be addressed to Nicholas A. Kotov, Qian Chen.

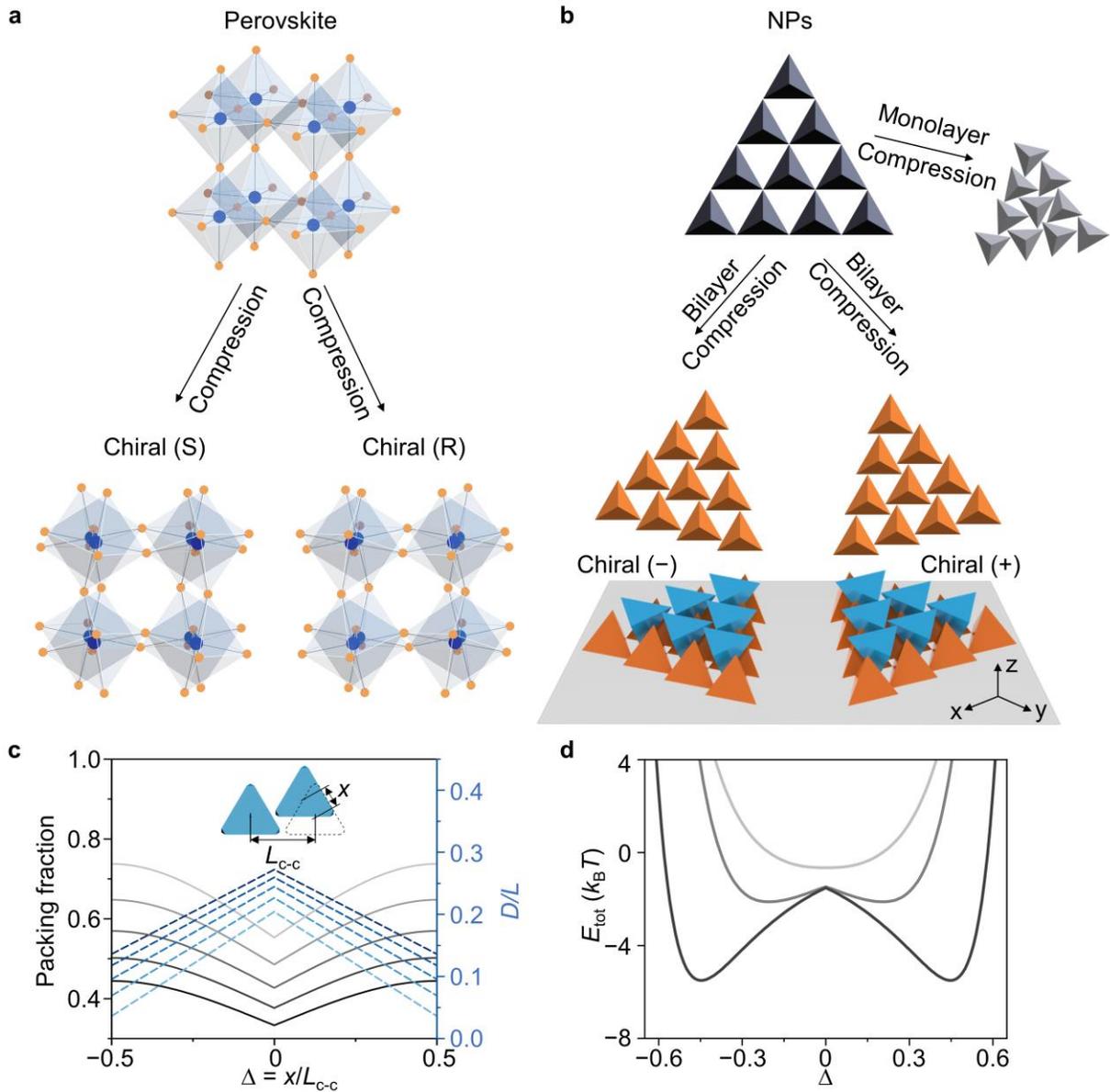

**Fig. 1. Chiral symmetry breaking in the superlattices from tetrahedra undergoing phase transition from low- to high-packing fraction states. a,** Schematics showing the transformation of non-closely packed achiral corner-sharing perovskite-like lattice into chiral states with higher packing fraction[20]. **b**, Schematics for chiral symmetry breaking of achiral corner-sharing bilayer lattice into pinwheel lattice on substrates. **c**, Packing fraction and $D$ changes as a function of corner offset $\Delta$, defined as the ratio of the travel distance $x$ and the initial centre-to-centre distance $L_{c-c}$ in the corner-to-corner alignment. From dark to light lines represent the change of the ratio of tetrahedron truncation (defined in Supplementary Note 1) from 0 to 0.16. **d**, Energy diagram showing the interparticle interaction as a function of $\Delta$. From light to dark lines represent the energy diagram for small, middle and large-sized tetrahedra, respectively. Calculations in **d** used the geometric parameters obtained from the experimental results (Supplementary Tables 3,4).

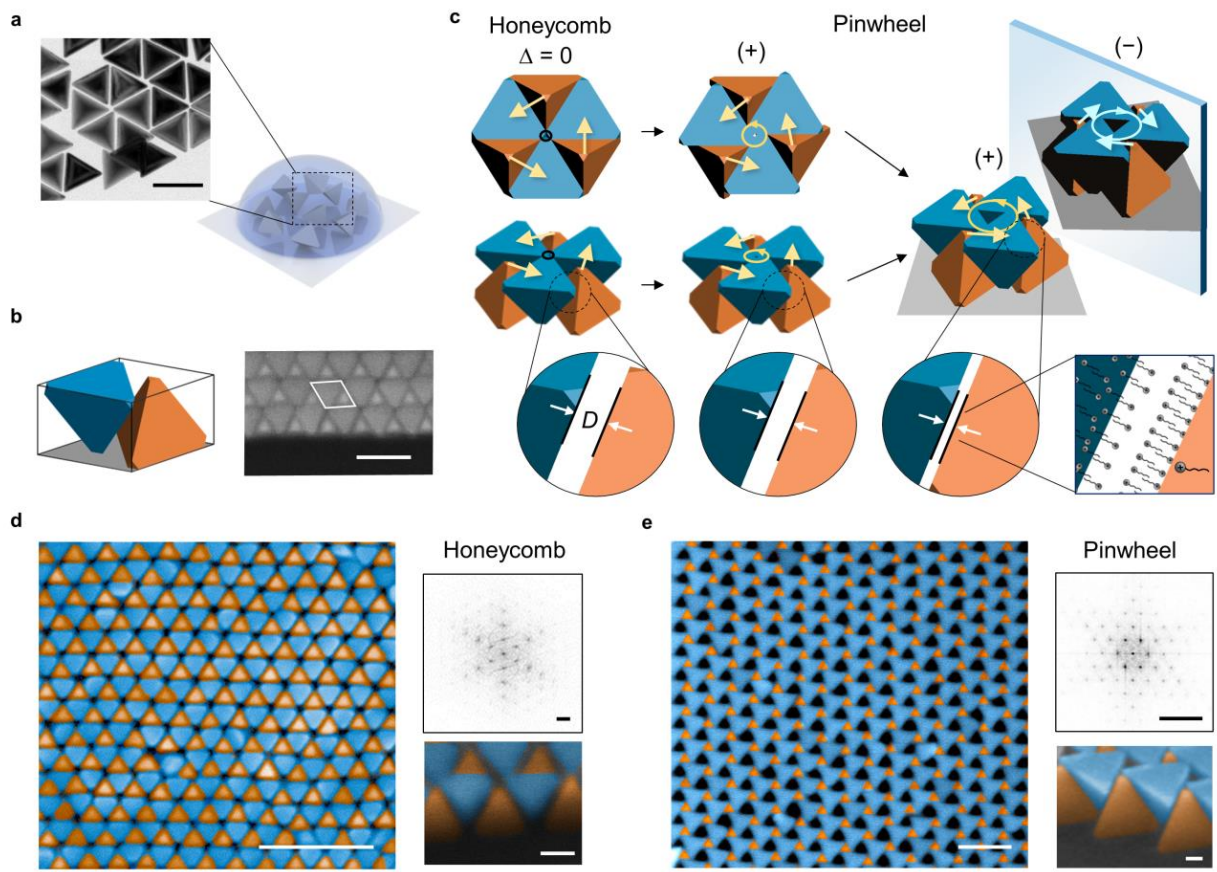

**Fig. 2. Corner-sharing and pinwheel bilayer superlattices with corner-to-edge connections from gold tetrahedral NPs. a**, Schematics of the evaporation-driven self-assembly on Si wafer and TEM image of gold tetrahedral building blocks. **b**, Schematics of the bilayer unit cell and SEM image showing the boundary of a corner-sharing bilayer domain. The white line in the SEM image showing the unit cell serves as the guide to eye. **c**, Schematics of the transition from corner-sharing tetrahedron bilayer lattice to pinwheel lattice. The twisting of tetrahedra in-plane is accompanied with decreasing face-to-face distance $D$ for interlayer neighbouring tetrahedra as shown in the zoomed-in schematics. Tetrahedra are coated by ligands as shown in the zoom-in schematics. **d,e**, SEM images (left), FFT (top right), and tilted SEM images (bottom right) of the large-scale extended corner-sharing lattice (**d**) and pinwheel lattice (**e**) formed by tetrahedral building blocks on substrates. Exact assembly conditions can be found in Supplementary Table 2. Scale bars: 100 nm (**a,b**); 200 nm (top view), 20 nm (tilted view) and 10 $nm^{-1}$ (FFT) (**d,e**).

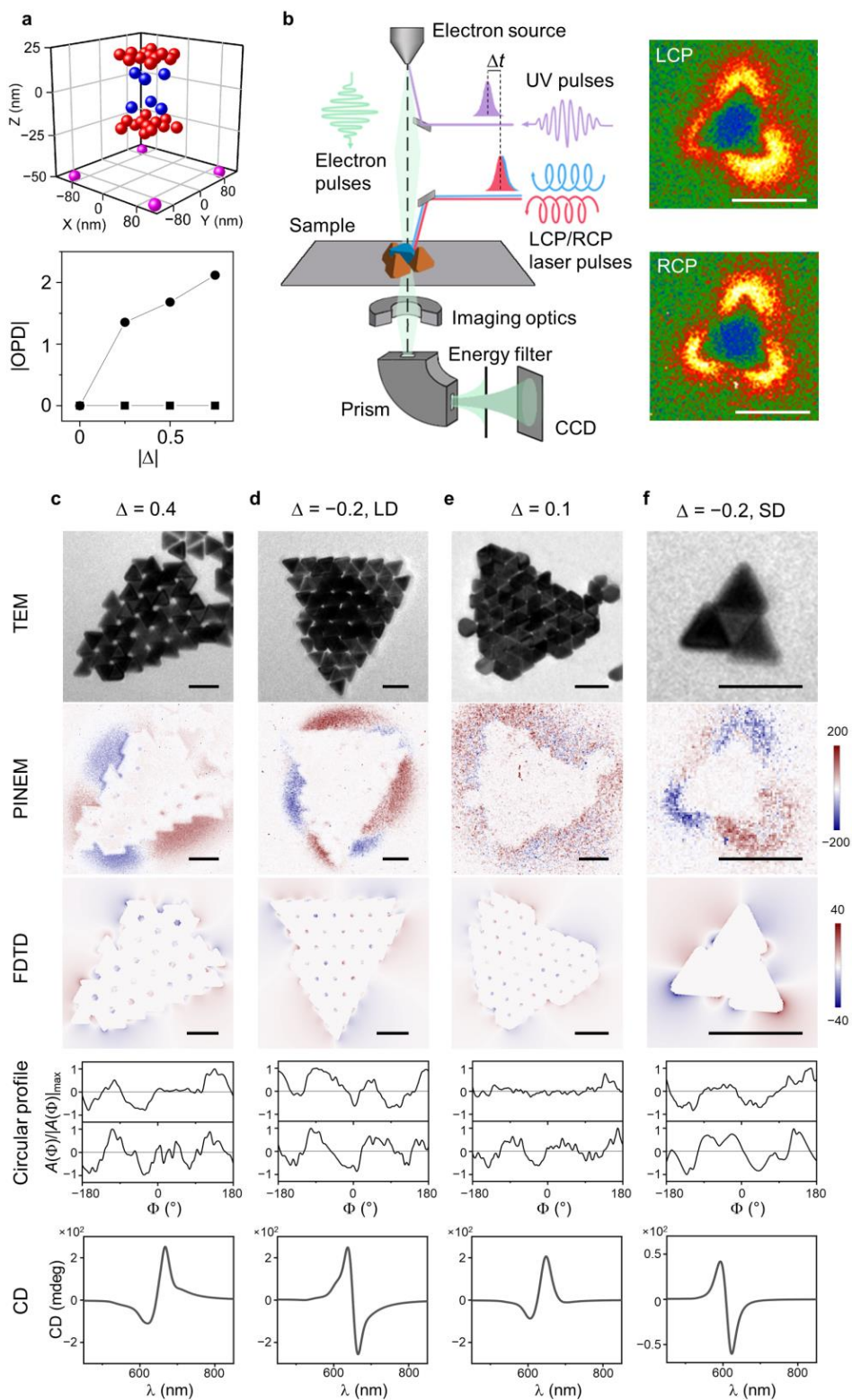

**Fig. 3. Chirality of pinwheel bilayer lattices deposited on solid substrates. a**, Osipov–Pickup–Dunmur (OPD) chirality measures calculated from extracted point groups in Cartesian coordinate

system (top) for different Δ (bottom) with (circular markers) and without (square markers) considering substrate. The point coordinates are extracted from corners (red dots) and CoMs (blue dots) of the pinwheel lattice with Δ = 0.5 and the substrate (magenta dots). **b**, Schematics of the PINEM setup for single NP-level chiroptical activity characterization. LCP or RCP laser pulses are applied as the excitation light source and photoexcite the plasmonic field. **c–f**, TEM images, PINEM subtracted *E*-field maps of LCP–RCP, FDTD simulated subtracted *E*-field distribution, normalized circular profiles, and CD spectra (from top to bottom) of chiral pinwheel bilayer lattices on TEM grids with different domain sizes as well as magnitudes and signs of Δ. SD stands for "small domain" and LD standards for "large domain" in (**d**,**f**). Normalized circular profiles are calculated based on *E*-field intensity $A$ as a function of rotation angle $\Phi$ and normalized by the absolute maximum *E*-field intensity $|A(\Phi)/_{max}$ for PINEM (top) and FDTD (bottom) subtracted maps. Scale bars: 100 nm.

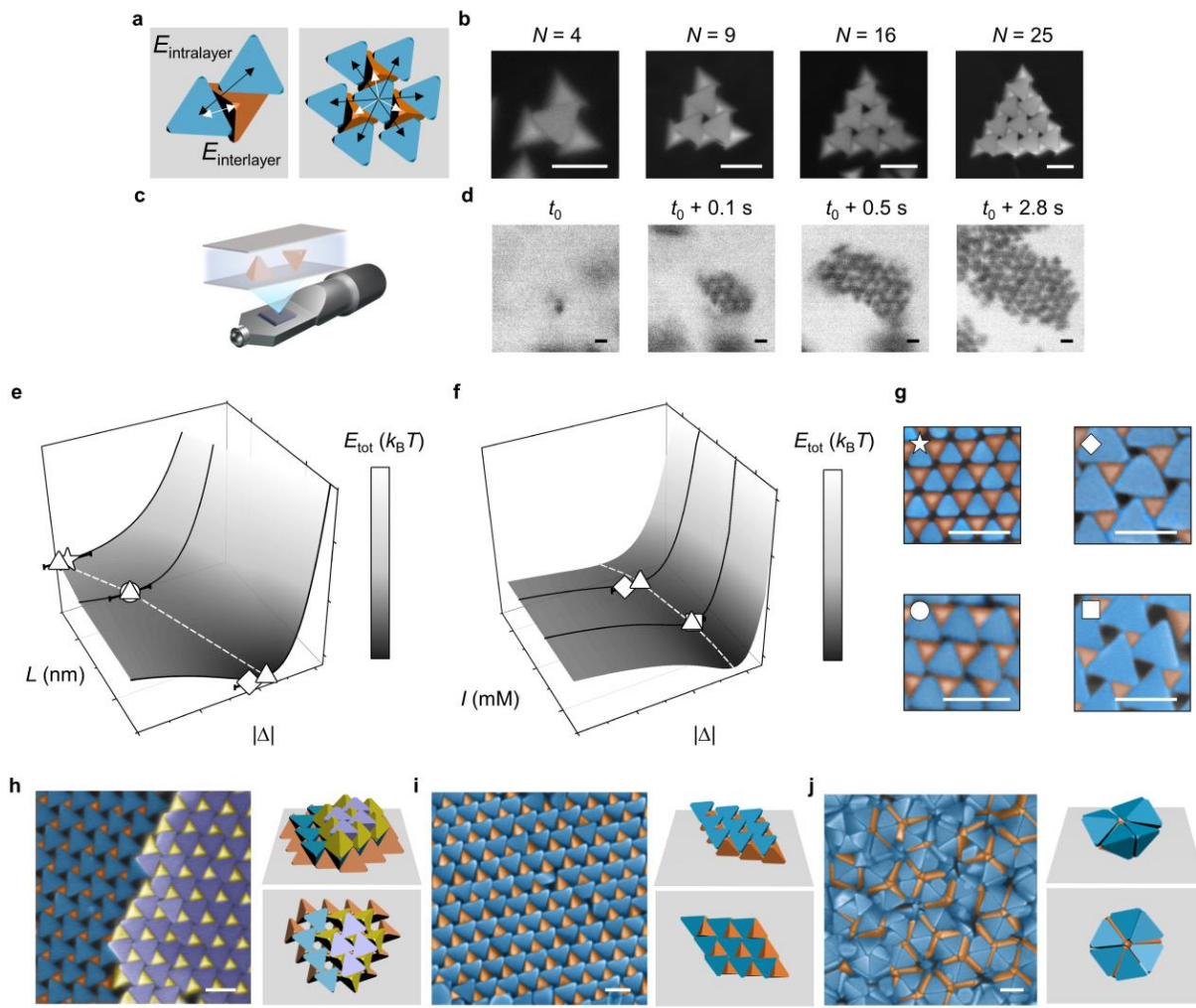

**Fig. 4. Formation mechanism and controllability of self-assembled chiral superlattices from tetrahedral NPs. a**, Models for interaction calculations considering one interlayer interactions and two intralayer interactions, as the representative motifs for chiral pinwheel lattice. **b**, SEM images of small bilayer domains formed at a lower particle concentration (Methods). **c,d**, Schematics (**c**) and snapshots (**d**) of in-situ nucleation and growth process of bilayer lattice using liquid-phase TEM. **e**, Three-dimensional (3D) energy diagram showing the interparticle interaction as a function of Δ and $L$. **f**, 3D energy diagram showing the interparticle interaction as a function of Δ and ionic strength $I$. For energy diagrams in **e** and **f**, star, circle, diamond and square markers with error bars indicate the experimental Δ and the corresponding standard deviations of the assembled structures shown in **g**, and triangles indicate the positions of energy minimum in the calculation given the same $L$ or $I$. **g**, SEM images of chiral bilayers assembled on substrate with differently sized tetrahedra and different initial salt concentration. **h–j**, SEM images and schematics of tetrahedra assembly driven by stronger interparticle attractions: multilayer with $z$-direction protrusion (**h**), tilted assembly without three-fold rotational symmetry (**i**), and face-to-face packing (**j**). Exact assembly conditions and lattice parameters can be found in Supplementary Tables 2–4. Scale bars: 100 nm.